\magnification \magstep1
\raggedbottom
\openup 2\jot
\voffset6truemm
\def\cstok#1{\leavevmode\thinspace\hbox{\vrule\vtop{\vbox{\hrule\kern1pt
\hbox{\vphantom{\tt/}\thinspace{\tt#1}\thinspace}}
\kern1pt\hrule}\vrule}\thinspace}
\centerline {\bf FOURTH-ORDER OPERATORS ON} 
\centerline {\bf MANIFOLDS WITH BOUNDARY}
\vskip 0.3cm
\leftline {Giampiero Esposito$^{1,2}$ and 
Alexander Yu. Kamenshchik$^{3,4}$}
\vskip 0.3cm
\noindent
{\it ${ }^{1}$Istituto Nazionale di Fisica Nucleare, Sezione di
Napoli, Mostra d'Oltremare Padiglione 20, 80125 Napoli, Italy}
\vskip 0.3cm
\noindent
{\it ${ }^{2}$Universit\`a di Napoli Federico II, Dipartimento
di Scienze Fisiche, Complesso Universitario di Monte S. Angelo,
Via Cintia, Edificio G, 80126 Napoli, Italy}
\vskip 0.3cm
\noindent
{\it ${ }^{3}$L. D. Landau Institute for Theoretical Physics
of Russian Academy of Sciences, Kosygina Str. 2, Moscow
117334, Russia}
\vskip 0.3cm
\noindent
{\it ${ }^{4}$Landau Network Coordination Centre, `A. Volta'
Centre for Scientific Culture, Villa Olmo, Via Cantoni 1,
22100 Como, Italy}
\vskip 1cm
\noindent
{\bf Abstract}. Recent work in the literature has studied
fourth-order elliptic operators on manifolds with boundary.
This paper proves that, in the case of the squared Laplace
operator, the boundary conditions which require that the
eigenfunctions and their normal derivative should vanish at
the boundary lead to self-adjointness of the 
boundary-value problem. On studying, for simplicity, the
squared Laplace operator in one dimension, on a closed interval
of the real line, alternative conditions which also ensure
self-adjointness set to zero at the boundary the eigenfunctions
and their second derivatives, or their first and third 
derivatives, or their second and third derivatives, or require
periodicity, i.e. a linear relation among the values of the
eigenfunctions at the ends of the interval. For the first
four choices of boundary conditions, the resulting one-loop
divergence is evaluated for a real scalar field on the portion
of flat Euclidean 4-space bounded by a 3-sphere, or by two
concentric 3-spheres.
\vskip 100cm
\leftline {\bf 1. Introduction}
\vskip 0.3cm
\noindent
The current attempts to develop quantum field theories of 
fundamental interactions have led to the consideration of
fourth-order or even higher-order differential operators on
closed Riemannian manifolds [1--5], or on manifolds with
boundary [6, 7]. The analysis of the transformation properties
under conformal rescalings of the background metric $g$
leads, in particular, to the consideration of conformally
covariant operators $P$, which transform according to the law
$$
P_{\omega}={\rm e}^{-(m+4)\omega /2}P(\omega=0)
{\rm e}^{(m-4)\omega /2}
\eqno (1.1)
$$
if $g$ rescales as $g_{\omega}={\rm e}^{2\omega}g$, $m$ being
the dimension of the Riemannian manifold which is studied. One
of the physical motivations for this analysis lies in the
possibility to use the Green functions of such operators to
build the effective action in curved space-times [5].

Another enlightening example is provided 
by the ghost sector of Euclidean Maxwell theory in
vacuum in four dimensions. The corresponding field equations are
well known to be invariant under conformal rescalings of $g$. 
On the other hand, the supplementary
(or gauge) conditions usually considered in the literature are
not invariant under conformal rescalings of $g$. Even just in
flat Euclidean 4-space, conformal invariance of the supplementary
condition is only achieved on making the 
Eastwood--Singer choice [8]:
$$
\nabla_{b}\nabla^{b}\nabla^{c}A_{c}=0
\eqno (1.2)
$$
where $A_{c}$ is the electromagnetic potential (a connection
1-form in geometric language). The preservation of Eq. (1.2) under
gauge transformations of $A_{c}$:
$$
{ }^{f}A_{c} \equiv A_{c}+\nabla_{c}f
\eqno (1.3)
$$
is achieved provided that $f$ obeys the fourth-order equation
$$
\cstok{\ }^{2}f=0
\eqno (1.4)
$$
where $\cstok{\ }^{2}$ is the box operator composed with itself:
$$
\cstok{\ }^{2} \equiv \nabla_{a}\nabla^{a}\nabla_{b}\nabla^{b}.
$$
In the corresponding quantum theory via path integrals, one thus
deals with two independent ghost fields (frequently referred to
as the ghost and the anti-ghost), both ruled by $\cstok{\ }^{2}$,
which is a fourth-order elliptic operator, and subject to the
following boundary conditions (hereafter, 
$\nabla_{N}\equiv N^{a}\nabla_{a}$ denotes
the covariant derivative along the inward-pointing normal
$N^{a}$ to the boundary):
$$
[\varepsilon]_{\partial M}=0
\eqno (1.5)
$$
$$
\Bigr[\nabla_{N}\varepsilon \Bigr]_{\partial M}=0 .
\eqno (1.6)
$$
Remarkably, since one now deals with a fourth-order elliptic
operator, it is insufficient to impose just Dirichlet or
Neumann (or Robin) boundary conditions. One needs instead both
(1.5) and (1.6), which are obtained from the following 
requirements: 
\vskip 0.3cm
\noindent
(i) Gauge invariance of the boundary conditions on $A_{b}$ [6, 7].
\vskip 0.3cm
\noindent
(ii) Conformal invariance of the whole set of boundary 
conditions [7].
\vskip 0.3cm
\noindent
(iii) Self-adjointness of the $\cstok{\ }^{2}$ operator 
(see section 2).
\vskip 0.3cm
\noindent
Although it remains extremely difficult to build a consistent 
quantization scheme via path-integral formalism for the full
Maxwell field in the Eastwood--Singer gauge (the gauge-field
operator on $A_{b}$ perturbations being, then, of sixth order
[6, 7]), the investigation of the ghost sector remains of considerable
interest in this case. There is in fact, on the one hand, the need
to understand how to quantize a gauge theory in a way which
preserves conformal invariance at all stages (as we just said),
and on the other hand the attempt to extend the recent work on 
conformally covariant operators [1--5] to the more realistic case 
of manifolds with boundary.

Our paper begins, therefore, with a detailed derivation of the
boundary conditions which ensure self-adjointness of the
operator ${d^{4}\over dx^{4}}$. For simplicity, the analysis
is limited to one-dimensional problems, but the key properties
are not affected by this sort of simplification. Section 3 proves
the strong ellipticity and self-adjointness of the resulting 
boundary value problem. Section 4
studies the one-loop properties of a scalar field on a portion
of flat Euclidean 4-space bounded by a 3-sphere, 
when the field is ruled by the
$\cstok{\ }^{2}$ operator and is subject to the boundary conditions
(1.5) and (1.6). Section 5 extends the analysis of section 4
to the part of flat Euclidean 4-space bounded by two concentric
3-spheres. Results and open problems are discussed in section 6,
and relevant details are described in the appendix.
\vskip 0.3cm
\leftline {\bf 2. Self-adjointness of the operator
${d^{4}\over dx^{4}}$}
\vskip 0.3cm
\noindent
We are concerned with the squared Laplace operator acting on
scalar fields on a flat Euclidean background, in the case when
curvature effects result from the boundary only. Moreover,
motivated by quantum cosmology and Euclidean quantum gravity,
the boundary is assumed to be a 3-sphere of radius $a$, 
or a pair of concentric 3-spheres [6, 7]. The former case, in
particular, may be viewed as the limiting case when the wave
function of the universe is studied at small 3-geometries
(i.e. as $a \rightarrow 0$), as shown in [9].

In our problem it is hence possible to expand
the scalar field on a family of 3-spheres centred on the origin,
according to the familiar relation [10]
$$
\varepsilon(x,\tau)=\sum_{n=1}^{\infty}\varepsilon_{n}(\tau)
Q^{(n)}(x) 
\eqno (2.1)
$$
where $\tau \in [0,a]$, $Q^{(n)}$ are the scalar harmonics on a
unit 3-sphere, $S^{3}$, and $x$ are local coordinates on
$S^{3}$. Thus, one is eventually led to study a one-dimensional
differential operator of fourth order, and this makes it clear 
why all the essential information is obtained by the analysis
of the operator $B \equiv {d^{4}\over dx^{4}}$ on a closed interval
of the real line, say $[0,1]$. The operator $B$ is required to act
on functions which are at least of class $C^{4}$ (see (2.22)),
and the following definition of scalar
product (anti-linear in the first argument) is considered:
$$
(u,v) \equiv \int_{0}^{1}u^{*}(x)v(x)dx .
\eqno (2.2)
$$
We now want to study under which conditions the operator $B$ is
self-adjoint, which means that it should be symmetric, and its
domain $D(B)$ should coincide with the domain of the adjoint 
$B^{\dagger}$. For this purpose, we first study the relation
between the scalar products $(Bu,v)$ and $(u,Bv)$. We have then
to integrate repeatedly by parts, using the Leibniz rule
to express
$$
{d\over dx}\left({d^{3}u^{*}\over dx^{3}}v \right), \;
{d\over dx}\left({d^{2}u^{*}\over dx^{2}}{dv\over dx}
\right), \;
{d\over dx}\left({du^{*}\over dx}{d^{2}v\over dx^{2}}
\right), \;
{d\over dx}\left(u^{*}{d^{3}v\over dx^{3}}\right) .
$$
This leads to
$$ \eqalignno{
(Bu,v)&=\left[{d^{3}u^{*}\over dx^{3}}v \right]_{0}^{1}
-\left[{d^{2}u^{*}\over dx^{2}}{dv\over dx}\right]_{0}^{1}
+\left[{du^{*}\over dx}{d^{2}v\over dx^{2}}\right]_{0}^{1} \cr
&-\left[u^{*}{d^{3}v \over dx^{3}}\right]_{0}^{1}
+(u,Bv) .
&(2.3)\cr}
$$
Bearing in mind that the adjoint, $B^{\dagger}$, of
${d^{4}\over dx^{4}}$ is again the operator 
${d^{4}\over dx^{4}}$, it is thus clear that the condition
$(Bu,v)=(u,B^{\dagger}v)$ is fulfilled provided that {\it both}
$u \in D(B)$ {\it and} $v \in D(B^{\dagger})$ obey the same
boundary conditions, for which the four terms expressing the
difference $(Bu,v)-(u,B^{\dagger}v)$ are found to vanish.
Some ways to achieve this are as follows.
\vskip 0.3cm
\noindent
(i) First option:
$$
u(0)=u(1)=0 \; \; \; \; u'(0)=u'(1)=0
\eqno (2.4)
$$
$$
v(0)=v(1)=0 \; \; \; \; v'(0)=v'(1)=0 .
\eqno (2.5)
$$
\vskip 0.3cm
\noindent
(ii) Second option:
$$
u(0)=u(1)=0 \; \; \; \; u''(0)=u''(1)=0
\eqno (2.6)
$$
$$
v(0)=v(1)=0 \; \; \; \; v''(0)=v''(1)=0 .
\eqno (2.7)
$$
\vskip 0.3cm
\noindent
(iii) Third option: 
$$
u'(0)=u'(1)=0 \; \; \; \; u'''(0)=u'''(1)=0
\eqno (2.8)
$$
$$
v'(0)=v'(1)=0 \; \; \; \; v'''(0)=v'''(1)=0.
\eqno (2.9)
$$
\vskip 0.3cm
\noindent
(iv) Fourth option:
$$
u''(0)=u''(1)=0 \; \; \; \; u'''(0)=u'''(1)=0
\eqno (2.10)
$$
$$
v''(0)=v''(1)=0 \; \; \; \; v'''(0)=v'''(1)=0 .
\eqno (2.11)
$$
\vskip 0.3cm
\noindent
(v) Periodic boundary conditions:
$$
{u(1)\over u(0)}=\beta
\eqno (2.12)
$$
$$
{u'(1)\over u'(0)}=\gamma
\eqno (2.13)
$$
$$
{u''(1)\over u''(0)}={1\over \gamma^{*}}
\eqno (2.14)
$$
$$
{u'''(1)\over u'''(0)}={1\over \beta^{*}}
\eqno (2.15)
$$
and the same for $v \in D(B^{\dagger})$, where $\beta$ and $\gamma$
are some constants, not necessarily equal. This is made possible by
the fourth-order nature of our operator. By contrast, if we were
studying the first-order 
operator $i{d\over dx}$ on the set of absolutely
continuous functions on $[0,1]$, periodic boundary conditions leading
to self-adjointness would involve one and the 
same complex parameter [11].

The solutions of the eigenvalue equation for the operator $B$, i.e.
$$
Bu \equiv {d^{4}u\over dx^{4}}=\lambda \; u
\eqno (2.16)
$$
read
$$
u(x)=C_{1} \cos \rho x + C_{2} \sin \rho x
+ C_{3} \cosh \rho x + C_{4} \sinh \rho x
\eqno (2.17)
$$
where $\rho \equiv \lambda^{1/4}$. In particular, the periodic
boundary conditions (2.12)--(2.15) lead to a linear algebraic
system for the evaluation of the coefficients $C_{1},C_{2},C_{3}$
and $C_{4}$ which admits non-trivial solutions if and only if the
determinant of the following matrix vanishes:
$$
\pmatrix{
(\cos \rho -\beta) & \sin \rho & (\cosh \rho - \beta)
& \sinh \rho \cr
-\sin \rho & (\cos \rho -\gamma) & \sinh \rho &
(\cosh \rho - \gamma) \cr
\left(-\cos \rho +{1\over \gamma^{*}}\right) & -\sin \rho &
\left(\cosh \rho -{1\over \gamma^{*}} \right) &
\sinh \rho \cr
\sin \rho & \left(-\cos \rho +{1\over \beta^{*}}\right)
& \sinh \rho & \left(\cosh \rho -{1\over \beta^{*}}
\right) \cr}.
$$
The above determinant, denoted by $\delta$, turns out to have the form
$$
\delta=F_{1}+F_{2}(\cos \rho +\cosh \rho)
+F_{3}(\cos \rho)(\cosh \rho)
\eqno (2.18)
$$
where
$$
F_{1} \equiv 4+2{\beta \over \beta^{*}}+2{\gamma \over \gamma^{*}}
+2{(2\beta \gamma+1)\over \beta^{*}\gamma^{*}}+2\beta \gamma
\eqno (2.19)
$$
$$
F_{2} \equiv - \left[2(\beta+\gamma)+2(\beta \gamma+1)
\left({1\over \beta^{*}}+{1\over \gamma^{*}}\right)
+2{(\beta+\gamma)\over \beta^{*}\gamma^{*}}\right]
\eqno (2.20)
$$
$$
F_{3} \equiv 2 \left[{(\beta+2\gamma)\over \beta^{*}}
+{(2\beta+\gamma)\over \gamma^{*}}+\beta \gamma
+{1\over \beta^{*}\gamma^{*}}\right].
\eqno (2.21)
$$

To sum up, if the conditions (2.4) and (2.5), or (2.6)
and (2.7), or (2.8) and (2.9), or (2.10) and (2.11),
or (2.12)--(2.15) are satisfied, the domains of
$B$ and of its adjoint turn out to coincide: 
$$ \eqalignno{
D(B)&=D(B^{\dagger}) \equiv
\left \{u: u \in AC^{4}[0,1], \;
(2.4) \; {\rm or} \; (2.6) \right . \cr
&\left . \; {\rm or} \; (2.8) \;
{\rm or} \; (2.10) \; {\rm or} \; 
(2.12)-(2.15) \; {\rm hold} \right \}.
&(2.22)\cr}
$$
With our notation, $AC^{4}[0,1]$ is the set of functions in
$L^{2}[0,1]$ whose weak derivatives up to third order are
absolutely continuous in [0,1], which ensures that the weak
derivatives, up to fourth order, are Lebesgue summable 
in [0,1], and that all $u$ in the domain are of class $C^{4}$
on [0,1]. Of course, symmetry of $B$ is also obtained with the 
boundary conditions just described.

In other words, at least five sets of boundary conditions, (i) or (ii) 
or (iii) or (iv) or (v), can be chosen
to ensure self-adjointness of the operator ${d^{4}\over dx^{4}}$.
Hereafter, we first consider the option (i), since, as was stated in the
introduction, it is the one which 
agrees with boundary conditions motivated by
the request of gauge invariance and conformal invariance, if the
scalar field is viewed as one of the two ghost fields of 
Euclidean Maxwell theory in the Eastwood--Singer gauge. We also
stress again that nothing is lost on studying just the 
``prototype" operator ${d^{4}\over dx^{4}}$. The one-dimensional
fourth-order operator may take a more complicated form in some
set of local coordinates (see (4.1)), but is always
reducible to the operator ${d^{4}\over dx^{4}}$ on the real
line (more precisely, a closed interval of $\Re$ in our problems).
\vskip 0.3cm
\leftline {\bf 3. Strong ellipticity and self-adjointness of the boundary 
value problem}
\vskip 0.3cm
\noindent
A number of points discussed in the previous section need to be
put on firmer ground. In particular, we are concerned with the
issue of ellipticity of the boundary value problem. This is
studied in terms of the leading symbol of our differential
operator, which is a squared Laplacian on a Riemannian manifold
with smooth boundary. It is indeed well known that the Fourier
transform makes it possible to associate to a differential
operator of order $k$ a polynomial of degree $k$, called 
characteristic polynomial or symbol. The leading symbol,
$\sigma_{L}$, picks out the highest order part of this polynomial.
For a squared Laplacian, denoted by $F$, it reads
$$
\sigma_{L}(F;x,\xi)=|\xi|^{4}I=g^{\mu \nu}g^{\rho \sigma}
\xi_{\mu}\xi_{\nu}\xi_{\rho}\xi_{\sigma}I.
\eqno (3.1)
$$
With a standard notation, $x$ are local coordinates on $M$, and
$\xi_{\mu}$ are cotangent vectors: $\xi_{\mu} \in T^{*}(M)$. The
leading symbol of $F$ is trivially elliptic in the interior of $M$,
since the right-hand side of (3.1) is positive-definite, and 
one has
$$
{\rm det} \Bigr(\sigma_{L}(F;x,\xi)-\lambda \Bigr)
=\Bigr(|\xi|^{4}-\lambda \Bigr)^{{\rm dim} \; V} \not = 0
\eqno (3.2)
$$
for all $\lambda \in {\bf C}-{\bf R}_{+}$, where $V$ is the vector
bundle over $M$ whose sections are the physical fields 
$\varphi$, acted upon by 
$F: C^{\infty}(V,M) \rightarrow C^{\infty}(V,M)$. In the presence
of a boundary, however, one needs a more careful definition of
ellipticity. First, for a manifold $M$ of dimension $m$, the $m$
coordinates $x$ are split into $m-1$ local coordinates on 
$\partial M$, denoted by $\left \{ {\hat x}^{k} \right \}$, and $r$, the
geodesic distance to the boundary. Similarly, 
the $m$ coordinates $\xi_{\mu}$ are
split into $m-1$ cotangent vectors $\zeta_{j} \in 
T^{*}(\partial M)$, jointly with a real parameter $\omega \in T^{*}(R)$.
The ellipticity we are interested in requires now that $\sigma_{L}$
should be elliptic in the interior of $M$, as specified before, and
that strong ellipticity should hold. This means that a unique solution
exists of the eigenvalue equation for the leading symbol:
$$
\Bigr[\sigma_{L}(F;\left \{ {\hat x}^{k} \right \},r=0,
\left \{ \zeta_{j} \right \}, \omega \rightarrow 
-i \partial_{r})- \lambda \Bigr]\varphi(r)=0
\eqno (3.3)
$$
subject to the boundary conditions and to a decay condition at infinity.
Before defining these concepts, note that, in (3.3), $i \omega$ is
eventually replaced by the operator of first derivative with respect to
the geodesic distance to the boundary.

A complete formulation of boundary conditions needs some abstraction. 
For this purpose, one has to consider two vector bundles, $W_{F}$
and $W_{F}^{'}$, over the boundary of $M$, with a {\it boundary
operator} $B_{F}$, relating their sections, i.e.
$$
B_{F}: C^{\infty}(W_{F},{\partial M}) \rightarrow
C^{\infty}(W_{F}^{'},{\partial M}).
$$
All the information about normal derivatives of the fields is not
encoded in $B_{F}$ but in the {\it boundary data} 
$\psi_{F}(\varphi) \in C^{\infty}(W_{F},{\partial M})$. For example,
with boundary conditions involving $\varphi$ and its first normal
derivative, one has
$$
\psi_{F}(\varphi)=\pmatrix{[\varphi]_{\partial M} \cr 
[\nabla_{N} \varphi]_{\partial M} \cr}
\eqno (3.4)
$$
$$
B_{F}=\pmatrix{I & 0 \cr 0 & I \cr}
\eqno (3.5)
$$
and the boundary conditions read $B_{F} \psi_{F}(\varphi)=0$. The
sections of the bundle $W_{F}^{'}$, which remained unspecified
so far, are obtained by applying to the sections of $W_{F}$ the
operator whose main diagonal coincides with the main diagonal of
$B_{F}$. More precisely, if the boundary conditions are mixed, 
on writing $B_{F}=P_{F}L$ for some
projector $P_{F}: W_{F} \rightarrow W_{F}^{'}$ and some operator
$$
L: C^{\infty}(W_{F},{\partial M}) \rightarrow 
C^{\infty}(W_{F},{\partial M})
$$
one has $\psi_{F}^{'} \in C^{\infty}(W_{F}^{'},{\partial M})$
realized as $\psi_{F}^{'}=P_{F}\chi$, for some 
$\chi \in C^{\infty}(W_{F},{\partial M})$. 
However, when the boundary
operator (3.5) is considered, the projector $P_{F}$ is turned 
into $B_{F}$, and the strong ellipticity condition demands that
a unique solution of Eq. (3.3) should exist, subject to the 
boundary condition
$$
\sigma_{g}(B_{F})(\left \{ {\hat x}^{k} \right \},
\left \{ \zeta_{j} \right \})\psi_{F}(\varphi)=\psi_{F}'(\varphi)
\; \; \; \forall \psi_{F}'(\varphi) 
\in C^{\infty}(W_{F}',{\partial M})
\eqno (3.6)
$$
and to the asymptotic condition
$$
\lim_{r \to \infty}\varphi(r)=0 .
\eqno (3.7)
$$
With a standard notation [7, 12], $\sigma_{g}(B_{F})$ is the 
{\it graded leading symbol} of the boundary operator $B_{F}$
in the local coordinates $\left \{ {\hat x}^{k} \right \},
\left \{ \zeta_{j} \right \}$. When $B_{F}$ takes the form (3.5),
$\sigma_{g}(B_{F})$ may be defined by
$$
\sigma_{g}(B_{F}) \equiv \pmatrix{I & 0 \cr 0 & I \cr}.
\eqno (3.8)
$$
Similarly to the case of the differential operator acting on
physical fields, one is here mapping the boundary operator into
its counterpart via Fourier transform. In the case of mixed
boundary conditions for operators of Laplace type [13], $B_{F}$
has off-diagonal elements which are first-order tangential 
operators, whereas complementary projectors occur on the main 
diagonal. One then finds a more elaborated structure [13]:
$$
\sigma_{g}(B_{F})=\pmatrix{\Pi & 0 \cr iT & I- \Pi \cr}
$$
where $T$ is an anti-self-adjoint matrix.

The asymptotic condition (3.7) picks out the solutions of the
eigenvalue equation (3.3) which satisfy (3.6) with
arbitrary boundary data $\psi_{F}'(\varphi)$ and vanish at infinite 
geodesic distance to the boundary. When all the above conditions
are satisfied $\forall \zeta \in T^{*}({\partial M}),
\forall \lambda \in {\bf C}-{\bf R}_{+}, \forall
(\zeta,\lambda) \not = (0,0)$ and $\forall \psi_{F}' \in
C^{\infty}(W_{F}^{'},{\partial M})$, one says that the boundary
value problem $(F,B_{F})$ for the squared Laplacian is strongly
elliptic with respect to the cone ${\bf C}-{\bf R}_{+}$. 
Following [14], we find the solution of (3.3), (3.6) {\it and} (3.7)
in the form ($\chi_{1}$ and $\chi_{2}$ being some constants)
$$
\varphi(r)=\chi_{1}{\rm e}^{-\rho_{1}r}
+\chi_{2}{\rm e}^{-\rho_{2}r}
\eqno (3.9)
$$
where, on setting
$$
|\zeta| \equiv \sqrt{g^{ij}({\hat x})\zeta_{i}\zeta_{j}}
\eqno (3.10)
$$
we define
$$
\rho_{1} \equiv + \sqrt{|\zeta|^{2}+\sqrt{\lambda}} 
\eqno (3.11)
$$
$$
\rho_{2} \equiv + \sqrt{|\zeta|^{2}-\sqrt{\lambda}}. 
\eqno (3.12)
$$
Comparison with the case of the Laplace operator shows that one
obtains strong ellipticity provided that $\pm \sqrt{\lambda} \in
{\bf C}-{\bf R}_{+}$, which yields [14]
$$
\lambda \in ({\bf C}-{\bf R}_{+}) \cap {\bf C}={\bf C}-{\bf R}_{+}.
\eqno (3.13)
$$
The boundary condition (3.6) leads to the equation $A \chi=\psi$,
where
$$
A \equiv \pmatrix{1&1 \cr -\rho_{1} & -\rho_{2} \cr}.
\eqno (3.14)
$$
This matrix is trivially invertible, and hence existence and
uniqueness of the solution is guaranteed. On writing
$\psi' \equiv \pmatrix{\psi_{0}^{'} \cr \psi_{1}^{'} \cr}$, where,
according to the rule described after (3.5), one has (for some
constants $\gamma_{1}$ and $\gamma_{2}$)
$$
\psi_{0}^{'}=\gamma_{1}+\gamma_{2}
\eqno (3.15)
$$
$$
\psi_{1}^{'}=-\rho_{1}\gamma_{1}-\rho_{2}\gamma_{2}
\eqno (3.16)
$$
one finds
$$
\chi_{1}=\gamma_{1}
\eqno (3.17)
$$
$$
\chi_{2}=\gamma_{2}.
\eqno (3.18)
$$

As stressed in [12], the condition of strong ellipticity is
{\it essential} to ensure the existence of the asymptotic expansions
normally assumed in the theory of heat-kernel asymptotics. In other
words, if one cannot prove strong ellipticity for a given choice of
boundary conditions, the local asymptotics of the fibre trace of the
heat-kernel diagonal, and the corresponding, global asymptotics
(resulting from integration over $M$) do not contain just the terms 
whose occurrence is ensured by invariance theory and by the Weyl
theorem on the invariants of the orthogonal group [12]. There are,
instead, highly singular contributions to the heat-kernel diagonal,
so that their integral over $M$ does not exist in any sense.

It is therefore reassuring to see that, precisely in the case more
relevant for quantum field theory [6, 7], strong ellipticity 
holds, and hence the resulting one-loop theory is well defined. The
following sections are devoted to a detailed evaluation of such
one-loop approximation, but now we should clarify the 
self-adjointness issue for the squared Laplacian. For this purpose, 
let us recall that, given a symmetric operator $A$, with
adjoint $A^{\dagger}$, its self-adjointness property can be studied
by evaluating the dimension of the space of solutions of the equation
$A^{\dagger}\varphi=i \varphi$, jointly with the corresponding
dimension for the equation $A^{\dagger}\varphi=-i \varphi$. More
precisely, one defines the {\it deficiency sub-spaces} [11]
$$
{\cal H}_{+} \equiv {\rm Ker}(i-A^{\dagger})
\eqno (3.19)
$$
$$
{\cal H}_{-} \equiv {\rm Ker}(i+A^{\dagger})
\eqno (3.20)
$$
and the associated {\it deficiency indices} [11]
$$
n_{+}(A) \equiv {\rm dim}[{\cal H}_{+}]
\eqno (3.21)
$$
$$
n_{-}(A) \equiv {\rm dim}[{\cal H}_{-}].
\eqno (3.22)
$$
Two theorems are then very useful [11]:
\vskip 0.3cm
\noindent
{\it Theorem 3.1.} Given a closed symmetric operator $A$ with deficiency
indices $n_{+}$ and $n_{-}$, $A$ is self-adjoint if and only if 
$n_{+}=0=n_{-}$. Moreover, $A$ has self-adjoint extensions if and
only if $n_{+}=n_{-}$, and a one-one correspondence exists between
self-adjoint extensions of $A$ and unitary maps from
${\cal H}_{+}$ onto ${\cal H}_{-}$.
\vskip 0.3cm
\noindent
{\it Theorem 3.2.} If $A$ is a symmetric operator with domain $D(A)$,
and if a conjugation $C$ exists which maps $D(A)$ into $D(A)$
and commutes with $A$: $CA=AC$, then $n_{+}(A)=n_{-}(A)$, and hence 
$A$ has self-adjoint extensions. 

In the case of the operator $B \equiv {d^{4}\over dx^{4}}$ studied
in section 2, since complex conjugation commutes with $B$, we
immediately know from theorem 3.2 that the deficiency indices of $B$
are equal. The solutions of the equations $B^{\dagger}\varphi=
\pm i \varphi$ on $L^{2}(0,\infty)$ (we shall later restrict to
[0,1]) are weak solutions. However, by virtue of the elliptic 
regularity theorem, these solutions are infinitely differentiable
and hence strong solutions [11]. Now the strong solutions of 
$$
{d^{4}\over dx^{4}}\varphi(x)=i \varphi(x)
\eqno (3.23)
$$
can be written in the form $\varphi(x)={\rm e}^{\alpha x}$, with
$\alpha$ a root of the equation $\alpha^{4}=i$. One then finds
$$
\alpha_{1}={\rm e}^{i {\pi \over 8}}=\cos {\pi \over 8}
+i \sin {\pi \over 8}
\eqno (3.24)
$$
$$
\alpha_{2}=i{\rm e}^{i{\pi \over 8}}=-\sin {\pi \over 8}
+i \cos {\pi \over 8}
\eqno (3.25)
$$
$$
\alpha_{3}=-{\rm e}^{i {\pi \over 8}}
=-\cos {\pi \over 8}-i \sin {\pi \over 8}
\eqno (3.26)
$$
$$
\alpha_{4}=-i {\rm e}^{i {\pi \over 8}}
=\sin {\pi \over 8}-i \cos {\pi \over 8} .
\eqno (3.27)
$$
Thus, only the strong solutions $\varphi_{2}(x) \equiv
{\rm e}^{\alpha_{2}x}$ and $\varphi_{3}(x) \equiv 
{\rm e}^{\alpha_{3}x}$ are in $L^{2}(0,\infty)$, and
$n_{+}(B)=2$. Similarly, the strong solutions of the equation
$$
{d^{4}\over dx^{4}}\varphi(x)=-i \varphi(x)
\eqno (3.28)
$$
can be written in the form $\varphi(x)={\rm e}^{\beta x}$, with
$\beta$ a root of the equation $\beta^{4}=-i$. One then finds
$$
\beta_{1}={\rm e}^{-i {\pi \over 8}}=\cos {\pi \over 8}
-i \sin {\pi \over 8}
\eqno (3.29)
$$
$$
\beta_{2}=i {\rm e}^{-i {\pi \over 8}}=\sin {\pi \over 8}
+i \cos {\pi \over 8}
\eqno (3.30)
$$
$$
\beta_{3}=-{\rm e}^{-i {\pi \over 8}}=-\cos {\pi \over 8}
+i \sin {\pi \over 8}
\eqno (3.31)
$$
$$
\beta_{4}=-i{\rm e}^{-i {\pi \over 8}}
=-\sin {\pi \over 8}-i \cos {\pi \over 8}
\eqno (3.32)
$$
which implies that only the strong solutions $\varphi_{3}(x)
\equiv {\rm e}^{\beta_{3} x}$ and $\varphi_{4}(x) \equiv
{\rm e}^{\beta_{4} x}$ are in $L^{2}(0,\infty)$, and hence
$n_{-}(B)=2=n_{+}(B)$. This property suggests also a non-trivial
link with the strong ellipticity analysis, where we have seen that
the asymptotic condition (3.7) selects only two of the original
four contributions to the solution (3.9).

However, on $L^{2}[0,1]$, all strong solutions resulting from
(3.24)--(3.27) and (3.29)--(3.32) are acceptable, and the
deficiency indices of the operator $B$ are, therefore, 
$n_{+}(B)=n_{-}(B)=4$. The domains for $B$ studied in section 2
and summarized in equation (2.22) are hence correctly interpreted
as domains of self-adjoint extensions of $B$.
\vskip 0.3cm
\leftline {\bf 4. One-loop divergence on the Euclidean 4-ball}
\vskip 0.3cm
\noindent
The definition and evaluation of functional determinants remains
a topic of crucial importance in quantum field theory. Here the
task is even more interesting, because we are studying a
fourth-order elliptic operator on a manifold with boundary. As
shown in [6, 7], the resulting eigenvalue equation for the 
modes occurring in the expansion (2.1) turns out to be, on the
Euclidean 4-ball,
$$ \eqalignno{
\; & \biggr[{d^{4}\over d\tau^{4}}+{6\over \tau}{d^{3}\over d\tau^{3}}
-{(2n^{2}-5)\over \tau^{2}}{d^{2}\over d\tau^{2}} \cr
&-{(2n^{2}+1)\over \tau^{3}}{d\over d\tau}
+{(n^{2}-1)^{2}\over \tau^{4}}\biggr]\varepsilon_{n}
=\lambda_{n}\varepsilon_{n}.
&(4.1)\cr}
$$
Thus, on setting $M \equiv \lambda_{n}^{1/4}$, the solution of
equation (4.1) is expressed by a linear combination of Bessel 
functions and modified Bessel functions [7], i.e.
$$ \eqalignno{
\varepsilon_{n}(\tau)&=A_{1,n}{I_{n}(M\tau)\over \tau}
+A_{2,n}{K_{n}(M\tau)\over \tau} \cr
&+A_{3,n}{J_{n}(M\tau)\over \tau}
+A_{4,n}{N_{n}(M\tau)\over \tau}.
&(4.2)\cr}
$$
Since the Euclidean 4-ball consists of a portion of flat
Euclidean 4-space bounded by a 3-sphere, the coefficients
$A_{2,n}$ and $A_{4,n}$ have to vanish $\forall n \geq 1$,
to ensure regularity of $\varepsilon_{n}$ 
at the origin. One is thus left with
scalar modes of the form
$$
\varepsilon_{n}(\tau)=A_{1,n}{I_{n}(M\tau)\over \tau}
+A_{3,n}{J_{n}(M\tau)\over \tau}.
\eqno (4.3)
$$
We focus on a $\zeta(0)$ calculation for such a set of massless
modes, subject to the boundary conditions (see (1.5), (1.6)
and (2.4), (2.5))
$$
[\varepsilon_{n}]_{\partial M}=0
\eqno (4.4)
$$
$$
[d \varepsilon_{n}/d\tau ]_{\partial M}=0
\eqno (4.5)
$$
because the resulting 1-loop analysis remains crucial in the
course of studying quantum theory as a theory of small disturbances
[15] of the underlying classical theory. Our calculation relies 
on the technique developed in [16] and applied several times
by the present authors (see [17] and references therein). 
The starting point is the remark that, since $\zeta$-functions
are $L^{2}$-traces of complex powers of elliptic operators, they
admit an integral representation with the help of the Cauchy
formula. For example, for a function $f$ analytic in the domain
bounded by a curve $\gamma$, one has
$$
\sum_{l=1}^{L}n_{l}f(z_{l})={1\over 2\pi i} \int_{\gamma}
f(z){d\over dz}\log F(z) \; dz
\eqno (4.6)
$$
where $F$ is a function having zeros at $z_{1},...,z_{L}$ with
multiplicities $n_{1},...,n_{L}$, respectively. Thus, on choosing
$f(z) \equiv z^{-s}$, one finds the desired integral representation
of the $\zeta$-function in the form
$$
\zeta(s)={1\over 2\pi i} \int_{\gamma} z^{-s} {d\over dz}
{\rm Tr} \log Q(z) \; dz
\eqno (4.7)
$$
where $\gamma$ is the contour in the complex-$z$ plane which
encircles all roots of the equation $Q(z)=0$, with $Q$ the function
expressing the equation obeyed by the eigenvalues by virtue of the
boundary conditions. The contour $\gamma$ is then deformed into a
new contour ${\widetilde \gamma}$, which encircles the cut in
the complex plane of the function $z^{-s}$, coinciding with the
negative real axis. After some technical steps, one 
eventually finds
$$
\zeta(s)={1\over 2\pi i} \int_{{\widetilde \gamma}} z^{-s}
{d\over dz}I(-z,s) \; dz
\eqno (4.8)
$$
where $I(-z,s)$ is the regularized infinite sum defined by
$I(-z,s) \equiv \sum_{(n)} n^{-2s} \log Q(z)$. 
More precisely, on denoting now by $f_{n}$ the function
occurring in the equation obeyed by the eigenvalues by virtue of
the boundary conditions, after taking out false roots
(e.g. $x=0$ is a false root of the equation 
$J_{\nu}(x)=0$), and
writing $d(n)$ for the degeneracy of the eigenvalues parametrized
by the integer $n$, one defines the function
$$
I(M^{2},s) \equiv \sum_{n=n_{0}}^{\infty}
d(n) n^{-2s}\log f_{n}(M^{2}).
\eqno (4.9)
$$
What is very useful is the analytic continuation
$``I(M^{2},s)"$ to the complex-$s$ plane of the function
$I(M^{2},s)$, which is a meromorphic function with a simple
pole at $s=0$, i.e.
$$
``I(M^{2},s)"={I_{\rm pole}(M^{2})\over s}
+I^{R}(M^{2})+{\rm O}(s) .
\eqno (4.10)
$$
The function $I_{\rm pole}$ is the residue at $s=0$, and makes
it possible to obtain the $\zeta(0)$ value as [16]
$$
\zeta(0)=I_{\rm log}+I_{\rm pole}(M^{2}=\infty)
-I_{\rm pole}(M^{2}=0)
\eqno (4.11)
$$
where $I_{\rm log}$ is the coefficient of the $\log(M)$ term
in $I^{R}$ as $M \rightarrow \infty$. The contributions 
$I_{\rm log}$ and $I_{\rm pole}(\infty)$ are obtained from the
uniform asymptotic expansions of basis functions as 
$M \rightarrow \infty$ and their order $n \rightarrow \infty$,
whilst $I_{\rm pole}(0)$ is obtained by taking the
$M \rightarrow 0$ limit of the eigenvalue condition, and then
studying the asymptotics as $n \rightarrow \infty$. More 
precisely, $I_{\rm pole}(\infty)$ coincides with the coefficient
of ${1\over n}$ in the asymptotic expansion as 
$n \rightarrow \infty$ of
$$
{1\over 2}d(n)\log[\rho_{\infty}(n)]
$$
where $\rho_{\infty}(n)$ is the $n$-dependent term in the 
eigenvalue condition as $M \rightarrow \infty$ and
$n \rightarrow \infty$. The $I_{\rm pole}(0)$ value is instead
obtained as the coefficient of ${1\over n}$ in the asymptotic
expansion as $n \rightarrow \infty$ of
$$
{1\over 2}d(n)\log[\rho_{0}(n)]
$$
where $\rho_{0}(n)$ is the $n$-dependent term in the eigenvalue
condition as $M \rightarrow 0$ and $n \rightarrow 
\infty$ [16, 17]. Although such a technique was originally developed
for second-order elliptic operators, it can be easily generalized
to study our fourth-order operators, provided that one bears in
mind that the eigenvalues have now dimension [length]$^{-4}$ (see
notation after (4.1)). Hence one should replace $M^{2}$ by $M^{4}$
in (4.9)--(4.11), but the final results remain unaffected.

In our problem, the equations (4.3)--(4.5) lead to the eigenvalue
condition (denoting by $a$ the radius of the 3-sphere)
$$
{\rm det} \pmatrix{I_{n}(Ma)& J_{n}(Ma) \cr
-I_{n}(Ma)+Ma I_{n}'(Ma) & 
-J_{n}(Ma)+Ma J_{n}'(Ma) \cr}=0
\eqno (4.12)
$$
which guarantees that non-trivial solutions exist for the
coefficients $A_{1,n}$ and $A_{3,n}$ in (4.3). At this stage, on
using the limiting form of Bessel functions $I_{n}$ and $J_{n}$
when the argument tends to zero (see (A.1)--(A.4)), 
one finds that the left-hand
side of (4.12) is proportional to $M^{2n}$ as 
$M \rightarrow 0$. Hence one has to multiply by $M^{-2n}$ to get
rid of false roots. Moreover, in the uniform asymptotic expansion 
of Bessel functions as $M \rightarrow \infty$ and 
$n \rightarrow \infty$, both $I$ and $J$ functions contribute a
${1\over \sqrt{M}}$ factor (see (A.5), (A.7) and (A.12)). 
These properties imply that $I_{\rm log}$ takes the value
$$
I_{\rm log}={1\over 2}\sum_{n=1}^{\infty}n^{2}(-2n)
=-\zeta_{R}(-3)=-{1\over 120}.
\eqno (4.13)
$$
The calculation of $I_{\rm pole}(\infty)$ relies on the
asymptotic expansions (A.5), (A.7) and (A.12) of the appendix.
One then finds that no $n$-dependent term occurs in the
eigenvalue condition (4.12), which implies
$$
I_{\rm pole}(\infty)=0 .
\eqno (4.14)
$$
Last, $I_{\rm pole}(0)$ is obtained after working out 
$\rho_{0}(n)$ for (4.12). For this purpose, we remark that,
as $M \rightarrow 0$ and $n \rightarrow \infty$, the first
line of the matrix in (4.12) consists of two elements both 
equal to ${1\over \Gamma(n+1)}$, whereas the second line
consists of two elements both equal to 
${(n-1)\over \Gamma(n+1)}$ (bearing in mind that all powers
of $(Ma/2)$ can be safely omitted, if one is interested in
$\rho_{0}(n)$). Hence one finds exact cancellation of the
contributions to $I_{\rm pole}(0)$, i.e.
$$
I_{\rm pole}(0)=0 .
\eqno (4.15)
$$
By virtue of (4.11) and (4.13)--(4.15) one finds, on the
Euclidean 4-ball, 
$$
\zeta(0)=-{1\over 120}
\eqno (4.16)
$$
for a real, massless scalar field.

It is interesting to notice that the effect of cancellation of
contributions to $I_{\rm pole}(0)$ arises because, whilst 
recurrence formulae for derivatives of Bessel functions:
$$
2J_{n}'(z)=J_{n-1}(z)-J_{n+1}(z)
\eqno (4.17)
$$
and those for modified Bessel functions:
$$
2I_{n}'(z)=I_{n-1}(z)+I_{n+1}(z)
\eqno (4.18)
$$
have different signs before $J_{n+1}(z)$ and $I_{n+1}(z)$,
respectively, only $J_{n-1}(z)$ and $I_{n-1}(z)$ survive in
the limit as $M \rightarrow 0$, and hence the determinant
(4.12) is equal to zero in that limit. 
On differentiating the recurrence
formulae (4.17) and (4.18) the appropriate number of times, 
one can easily check that the corresponding determinants for
the boundary conditions (ii), (iii) and (iv) of section 2 are
also equal to zero at $M=0$. Thus, the contributions to
$I_{\rm pole}(0)$ with these boundary conditions vanish as well.

Moreover, the expression (4.13) for $I_{\rm log}$ is trivially
modified by adding to $(-2n)$ the integer numbers 1 for (ii),
3 for (iii) and 4 for (iv), respectively. Their contributions
are proportional to
$$
\sum_{n=1}^{\infty}n^{2}=\zeta_{R}(-2)=0.
$$
Thus, the results (4.13) and (4.16) hold for the first four types 
of boundary conditions.
\vskip 0.3cm
\leftline {\bf 5. One-loop divergence in the two-boundary problem}
\vskip 0.3cm
\noindent
In the two-boundary problem one studies a portion of flat
Euclidean 4-space bounded by two concentric 3-spheres. This case
is very interesting because it is more directly related to the
familiar framework in quantum field theory, where one normally
assigns boundary data on two three-surfaces (it should be
stressed, however, that unlike scattering problems we are
considering a path-integral representation of amplitudes in
a finite region).

On denoting by $a$ and $b$, with $a > b$, the radii of the 
two concentric 3-sphere boundaries, we can consider the
complete form (4.2) of our scalar modes, because no singularity
at the origin occurs in the two-boundary problem, and hence all
linearly independent integrals are regular, for all 
$\tau \in [b,a]$. We now impose the boundary conditions (1.5)
and (1.6), which lead to the eigenvalue condition
$$
{\rm det} \pmatrix{I_{n}(Mb) & K_{n}(Mb) & J_{n}(Mb)
& N_{n}(Mb) \cr
F_{I_{n}}(Mb) & F_{K_{n}}(Mb) & F_{J_{n}}(Mb)
& F_{N_{n}}(Mb) \cr
I_{n}(Ma) & K_{n}(Ma) & J_{n}(Ma) & N_{n}(Ma) \cr
F_{I_{n}}(Ma) & F_{K_{n}}(Ma) & F_{J_{n}}(Ma)
& F_{N_{n}}(Ma) \cr}=0
\eqno (5.1)
$$
where, for $Z=I,K,J$ or $N$, we define
$$
F_{Z_{n}}(Mx) \equiv -Z_{n}(Mx)+Mx Z_{n}'(Mx) .
\eqno (5.2)
$$
On using the approximate forms (A.1)--(A.4) of Bessel functions,
one finds that the left-hand side of (5.1) is proportional
to $M^{0}$ as $M \rightarrow 0$. Hence there are no
false roots of (5.1).
As a next step we notice that, as $M \rightarrow
\infty$ and $n \rightarrow \infty$, the $I_{n},K_{n},J_{n}$
and $N_{n}$ functions contribute a ${1\over \sqrt{M}}$ factor,
whereas $F_{Z_{n}}$, defined in (5.2), contributes a
$\sqrt{M}$ factor. Moreover, the dominant contribution to (5.1)
as $M \rightarrow \infty$ and $n \rightarrow \infty$ is given by
$K_{n}(Mb), N_{n}(Mb), F_{K_{n}}(Mb), F_{N_{n}}(Mb)$ (in the
first two rows), jointly with $I_{n}(Ma), J_{n}(Ma), 
F_{I_{n}}(Ma), F_{J_{n}}(Ma)$ (in the last two rows). One then 
finds that $I_{\rm log}$ vanishes, because 
$$
I_{\rm log}={1\over 2}\sum_{n=1}^{\infty}n^{2} \cdot 0=0.
\eqno (5.3)
$$
The value of $I_{\rm pole}(\infty)$ also vanishes, because no
$n$-dependent term occurs in equation (5.1) when
$n \rightarrow \infty$ and $M \rightarrow \infty$. Moreover,
$I_{\rm pole}(0)$ vanishes as well, since the determinant 
leading to $\rho_{0}(n)$ takes the form (by virtue of
(A.1)--(A.4))
$$
D(n)={\rm det} \pmatrix{0& {1\over 2}\Gamma(n) & 0
& -{1\over \pi}\Gamma(n) \cr
0 & -{1\over 2}\Gamma (n+1) & 0 & {1\over \pi}
\Gamma(n) (n+1) \cr
{1\over \Gamma(n+1)} & 0 & {1\over \Gamma(n+1)} & 0 \cr
{1\over \Gamma(n+1)} & 0 & {1\over \Gamma(n+1)} & 0 \cr}
\eqno (5.4)
$$
and this vanishes exactly. 

To sum up, we find that the $\zeta(0)$ value is zero
in the two-boundary problem:
$$
\zeta(0)= 0 .
\eqno (5.5)
$$
With the same arguments presented in the end of section 4, the
result (5.5) is found to hold for the first four boundary conditions
studied in section 2.
\vskip 10cm
\leftline {\bf 6. Concluding remarks}
\vskip 0.3cm
\noindent
The analysis of the squared Laplace operator in flat Euclidean
backgrounds is motivated by the ghost sector of Euclidean
Maxwell theory in a conformally invariant gauge, but has been
here restricted to a real scalar field. Further motivations
result from the theory of conformally covariant operators, which
is an important branch of spectral geometry, and finds applications
also in Euclidean quantum gravity [1--7]. The  
contributions of our investigation are as follows.
\vskip 0.3cm
\noindent 
(i) The boundary conditions for which the squared Laplace 
operator is self-adjoint have been derived (cf [18]), 
taking as prototype the operator ${d^{4}\over dx^{4}}$ on a
closed interval of the real line. 
Interestingly, at least five sets of
boundary conditions are then found to arise, and the option
described by (2.4) and (2.5) coincides, if the field in (2.1)
were a ghost field, with the boundary conditions obtained from
the request of gauge invariance of the boundary conditions on 
$A_{b}$, when the Eastwood--Singer supplementary condition
is imposed. The general reader, however, should be aware that 
the above boundary conditions have already been studied in the
mathematical literature. For example, Eq. (1.5.51) of [18]
studies the more involved boundary value problem
$$
\Bigr(\bigtriangleup^{2}+\lambda \bigtriangleup \Bigr)v=f
\; \; {\rm in} \; \; \Omega
\eqno (6.1)
$$
$$
v=0 \; \; {\rm at} \; \; {\partial \Omega}
\eqno (6.2)
$$
$$
\nabla_{N}v=\Bigr(1-\bigtriangleup_{\Gamma}\Bigr)^{-{1\over 2}}
\varphi \; \; {\rm at} \; \; {\partial \Omega}
\eqno (6.3)
$$
where $\bigtriangleup_{\Gamma}$ is chosen in such a way that 
$\Bigr(1-\bigtriangleup_{\Gamma}\Bigr)^{1\over 2}$ is a
suitable bijective operator [18]. Here it is enough to remark
that, since $\bigtriangleup^{2}+\lambda \bigtriangleup
\geq \bigtriangleup^{2}$ for all $\lambda \in 
{\overline {\bf R}}_{-}$, the solution of (6.1)--(6.3) is unique
for all smooth data. Hence strong ellipticity follows [18].
\vskip 0.3cm
\noindent
(ii) Given the fourth-order eigenvalue equation (4.1), the
contribution of the corresponding eigenmodes to the one-loop
divergence has been derived for the first time on the 
Euclidean 4-ball (see (4.16)), or on the portion of flat 
Euclidean 4-space bounded by two concentric 3-spheres
(see (5.5)).
\vskip 0.3cm
\noindent
In our opinion, the property (i) is crucial because no complete
prescription for the quantization is obtained unless suitable
sets of boundary conditions are imposed, and sections 2 and 3
represent a non-trivial step in this direction.
The result (ii) is instead relevant for the analysis of
one-loop semiclassical effects in quantum field theory. In other
words, if one has to come to terms with higher order differential
operators in the quantization of gauge theories and gravitation,
it appears necessary to develop techniques for a systematic
investigation of one-loop ultraviolet divergences, as a
first step towards a thorough understanding of 
their perturbative properties.

Some outstanding problems are now in sight. First, it appears
interesting to extend our mode-by-mode analysis to curved
backgrounds with boundary. In this case, the fourth-order
conformally covariant differential operator is more complicated
than the squared Laplace operator, and involves also the
Ricci curvature and the scalar curvature of the background.
Second, one should use Weyl's theorem on the invariants of
the orthogonal group to understand the general structure of
heat-kernel asymptotics [12] for fourth-order differential
operators on manifolds with boundary. A naturally occurring 
question within that framework is, to what extent functorial 
methods [7, 12] can then be used to compute all heat-kernel
coefficients for a given form of the differential operator
and of the boundary operator. Third, the recently considered
effect of tangential derivatives in the boundary operator 
[13, 19--21] might give rise to generalized boundary conditions
for conformally covariant operators. The appropriate 
mathematical theory is still lacking in the literature, but
would be of much help for the current attempts to understand
the formulation of quantum field theories on manifolds with
boundary [7, 13, 17, 19--21].
\vskip 0.3cm
\leftline {\bf Acknowledgments}
\vskip 0.3cm
\noindent 
This work has been partially supported by PRIN97 `Sintesi'.
A K was partially supported by RFBR via Grant No. 
96-02-16220, and by the RFBR Grant for support of leading
scientific schools No. 96-15-96458. 
A K is also grateful to the CARIPLO Scientific Foundation
for financial support.
\vskip 0.3cm
\leftline {\bf Appendix} 
\vskip 0.3cm
\noindent
In sections 4 and 5 we need the asymptotic expansions of Bessel
functions when the argument tends to zero, or when both the
argument and the order are very large [22]. In the former case, 
one finds, for all $n \geq 1$,
$$
I_{n}(x) \sim {(x/2)^{n}\over \Gamma(n+1)}
\eqno ({\rm A}.1)
$$
$$
J_{n}(x) \sim {(x/2)^{n}\over \Gamma(n+1)}
\eqno ({\rm A}.2)
$$
$$
K_{n}(x) \sim {1\over 2}\Gamma(n)(x/2)^{-n}
\eqno ({\rm A}.3)
$$
$$
N_{n}(x) \sim -{1\over \pi}\Gamma(n)(x/2)^{-n}.
\eqno ({\rm A}.4)
$$
Moreover, when the argument is greater than the order, both
being large, one finds, for modified Bessel functions [23],
$$
I_{n}(nz) \sim {{\rm e}^{ny}\over \sqrt{2\pi n}
(1+z^{2})^{1\over 4}} \sum_{s=0}^{\infty}{U_{s}(y)\over n^{s}}
\eqno ({\rm A}.5)
$$
$$
K_{n}(nz) \sim \sqrt{\pi \over 2n}
{{\rm e}^{-ny}\over (1+z^{2})^{1\over 4}}
\sum_{s=0}^{\infty}(-1)^{s}{U_{s}(y)\over n^{s}}
\eqno ({\rm A}.6)
$$
$$
I_{n}'(nz) \sim {(1+z^{2})^{1\over 4}\over z}
{{\rm e}^{ny}\over \sqrt{2\pi n}}
\sum_{s=0}^{\infty}{V_{s}(y)\over n^{s}}
\eqno ({\rm A}.7)
$$
$$
K_{n}'(nz) \sim -\sqrt{\pi \over 2n}
{(1+z^{2})^{1\over 4}\over z}{\rm e}^{-ny}
\sum_{s=0}^{\infty}(-1)^{s}{V_{s}(y)\over n^{s}}
\eqno ({\rm A}.8)
$$
where
$$
y \equiv \sqrt{1+z^{2}} + \log {z\over 1+\sqrt{1+z^{2}} } .
\eqno ({\rm A}.9)
$$
We do not need, in our calculations, the detailed form of the
polynomials $U_{s}$ and $V_{s}$, which are generated by
recurrence relations and can be found in [23]. For the Bessel
functions $J_{n}(nz)$ and $N_{n}(nz)$, with both $n$ and $z$
very large, we need the asymptotic expansions in section 8.41
of [22], which are more conveniently expressed after setting
$z \equiv \sec \beta$:
$$ \eqalignno{
\; & J_{n}(n \sec \beta) \sim \sqrt{2\over n \pi \tan \beta}
\left[\cos \left (n \tan \beta -n \beta -{\pi\over 4} \right)
\sum_{m=0}^{\infty}{(-1)^{m}\Gamma \left(2m+{1\over 2} \right)
\over \Gamma \left({1\over 2}\right)}
{A_{2m}\over {\left({n\over 2}\tan \beta \right)}^{2m}}
\right . \cr
& \left . + \sin \left(n \tan \beta -n \beta 
-{\pi \over 4} \right)
\sum_{m=0}^{\infty}{(-1)^{m}\Gamma \left(2m+{3\over 2} \right)
\over \Gamma \left({1\over 2}\right)}
{A_{2m+1}\over {\left({n\over 2}\tan \beta \right)}^{2m+1}}
\right] &({\rm A}.10)\cr}
$$
$$ \eqalignno{
\; & N_{n}(n \sec \beta) \sim \sqrt{2\over n \pi \tan \beta}
\left[\sin \left (n \tan \beta -n \beta -{\pi\over 4} \right)
\sum_{m=0}^{\infty}{(-1)^{m}\Gamma \left(2m+{1\over 2} \right)
\over \Gamma \left({1\over 2}\right)}
{A_{2m}\over {\left({n\over 2}\tan \beta \right)}^{2m}}
\right . \cr
& \left . - \cos \left(n \tan \beta -n \beta 
-{\pi \over 4} \right)
\sum_{m=0}^{\infty}{(-1)^{m}\Gamma \left(2m+{3\over 2} \right)
\over \Gamma \left({1\over 2}\right)}
{A_{2m+1}\over {\left({n\over 2}\tan \beta \right)}^{2m+1}}
\right] &({\rm A}.11)\cr}
$$
where $A_{2m}$ and $A_{2m+1}$ are numerical coefficients.
In particular, it is crucial to pick out the dominant terms of 
the expansions (A.10) and (A.11), i.e.
$$ \eqalignno{
J_{n}(nz) & \sim \sqrt{2\over n \pi}(z^{2}-1)^{-{1\over 4}}
\biggr[\cos \left(n \left(\sqrt{z^{2}-1}-\arccos {1\over z}
\right)-{\pi \over 4}\right) \cr
&+{\rm O}(n^{-1})\biggr]
&({\rm A}.12)\cr}
$$
$$ \eqalignno{
N_{n}(nz) & \sim \sqrt{2\over n \pi}(z^{2}-1)^{-{1\over 4}}
\biggr[\sin \left(n \left(\sqrt{z^{2}-1}-\arccos {1\over z}
\right)-{\pi \over 4}\right) \cr
&+{\rm O}(n^{-1})\biggr].
&({\rm A}.13)\cr}
$$
Thus, in (4.12), both $J_{n}(Ma)$ and $J_{n}'(Ma)$ contribute
${1\over \sqrt{M}}$ at large $M$ (as well as $I_{n}(Ma)$ 
and $I_{n}'(Ma)$). In section 5, all Bessel functions and 
their first derivatives contribute, for the same reason,
a factor ${1\over \sqrt{M}}$ in the eigenvalue condition
when both $n$ and $M$ tend to $\infty$.
\vskip 5cm
\leftline {\bf References}
\vskip 0.3cm
\noindent
\item {[1]}
Branson T P 1996 {\it Commun. Math. Phys.} {\bf 178} 301
\item {[2]}
Avramidi I G 1997 {\it Phys. Lett.} {\bf 403B} 280
\item {[3]}
Avramidi I G 1998 {\it J. Math. Phys.} {\bf 39} 2889
\item {[4]}
Erdmenger J 1997 {\it Class. Quantum Grav.} {\bf 14} 2061
\item {[5]}
Erdmenger J and Osborn H 1998 {\it Class. Quantum Grav.}
{\bf 15} 273
\item {[6]}
Esposito G 1997 {\it Phys. Rev.} D {\bf 56} 2442
\item {[7]}
Esposito G 1998 {\it Dirac Operators and Spectral Geometry}
({\it Cambridge Lecture Notes in Physics 12})
(Cambridge: Cambridge University Press)
\item {[8]}
Eastwood M and Singer I M 1985 {\it Phys. Lett.} 
{\bf 107A} 73
\item {[9]}
Schleich K 1985 {\it Phys. Rev.} D {\bf 32} 1889
\item {[10]}
Lifshitz E M and Khalatnikov I M 1963 {\it Adv. Phys.}
{\bf 12} 185
\item {[11]}
Reed M and Simon B 1975 {\it Methods of Modern Mathematical Physics.
II. Fourier Analysis and Self-Adjointness} (New York: Academic)
\item {[12]}
Gilkey P B 1995 {\it Invariance Theory, the Heat Equation and
the Atiyah-Singer Index Theorem} (Boca Raton: Chemical Rubber
Company)
\item {[13]}
Avramidi I G and Esposito G 1997 Gauge theories on manifolds
with boundary {\it Preprint} hep-th/9710048
\item {[14]}
Esposito G 1998 Boundary value problems for the squared 
Laplace operator {\it Preprint} hep-th/9809031
\item {[15]}
DeWitt B S 1965 {\it Dynamical Theory of Groups and Fields}
(New York: Gordon and Breach)
\item {[16]}
Barvinsky A O, Kamenshchik A Yu and Karmazin I P 1992
{\it Ann. Phys. (N.Y.)} {\bf 219} 201
\item {[17]}
Esposito G, Kamenshchik A Yu and Pollifrone G 1997 
{\it Euclidean Quantum Gravity on Manifolds with Boundary}
({\it Fundamental Theories of Physics 85}) (Dordrecht: Kluwer)
\item {[18]}
Grubb G 1996 {\it Func\-tio\-nal Cal\-cu\-lus 
of Pse\-u\-do-Dif\-fe\-ren\-ti\-al
Bo\-un\-da\-ry Pro\-ble\-ms} ({\it Pro\-gre\-ss 
in Ma\-the\-ma\-ti\-cs 65})
(Bo\-ston: Birkh\"{a}user)
\item {[19]}
McAvity D M and Osborn H 1991 {\it Class. Quantum Grav.} 
{\bf 8} 1445
\item {[20]}
Avramidi I G and Esposito G 1998 {\it Class. Quantum Grav.}
{\bf 15} 281
\item {[21]}
Dowker J S and Kirsten K 1997 {\it Class. Quantum Grav.}
{\bf 14} L169
\item {[22]}
Watson G N 1966 {\it A Treatise on the Theory of Bessel Functions}
(Cambridge: Cambridge University Press)
\item {[23]}
Olver F W J 1954 {\it Philos. Trans. Roy. Soc. London}
{\bf A 247} 328

\bye